\documentclass[twocolumn,pre,floatfix,superscriptaddress]{revtex4}

\usepackage{color}
\usepackage{tabularx}
\usepackage{makecell}
\usepackage{dblfloatfix}
\usepackage{fixltx2e}
\usepackage{epsfig}
\usepackage{amsmath}
\usepackage{soul}
\usepackage{amssymb}
\usepackage{bm}
\usepackage{graphicx}
\usepackage{wasysym}
\usepackage{enumitem}
\usepackage{epsfig}
\usepackage{amsmath}
\usepackage[svgnames]{xcolor}
\usepackage[pdftex]{hyperref}
\usepackage{times}
\hypersetup{colorlinks=true,
            linkcolor=blue,
            citecolor=ForestGreen,
            urlcolor=DarkOrchid
}
\usepackage{etoolbox}
\makeatletter
\patchcmd{\math@cr@@@align}% <cmd>
  {\place@tag}% <search>
  {\bgroup\color{black}\place@tag\egroup}% <replace>
  {}{}% <success><failure>
\makeatother

\begin{document}

\title{Numerical observation of a glassy phase in the three-dimensional Coulomb glass}

\author{Amin Barzegar}
\affiliation{Department of Physics and Astronomy, Texas A\&M University,
College Station, Texas 77843-4242, USA}

\author{Juan Carlos Andresen}
\affiliation{Department of Physics, Ben Gurion University of the Negev, Beer Sheva 84105, Israel}

\author{Moshe Schechter}
\affiliation{Department of Physics, Ben Gurion University of the Negev, Beer Sheva 84105, Israel}

\author{Helmut G. Katzgraber}
\affiliation{Microsoft Quantum, Microsoft, Redmond, Washington 98052, USA}
\affiliation{Department of Physics and Astronomy, Texas A\&M University, College Station, Texas 77843-4242, USA}
\affiliation{Santa Fe Institute, Santa Fe, New Mexico 87501 USA}

\begin{abstract}

The existence of an equilibrium glassy phase for charges in a disordered
potential with long-range electrostatic interactions has remained
controversial for many years. Here we conduct an extensive numerical
study of the disorder–temperature phase diagram of the
three-dimensional Coulomb glass model using population annealing Monte
Carlo to thermalize the system down to extremely low temperatures. Our
results strongly suggest that, in addition to a charge order phase, a
transition to a glassy phase can be observed, consistent with previous
analytical and experimental studies.

\end{abstract}
\date{\today}
\maketitle

\section{Introduction}
\begin{figure}[t!]
\includegraphics[width=0.9\columnwidth]{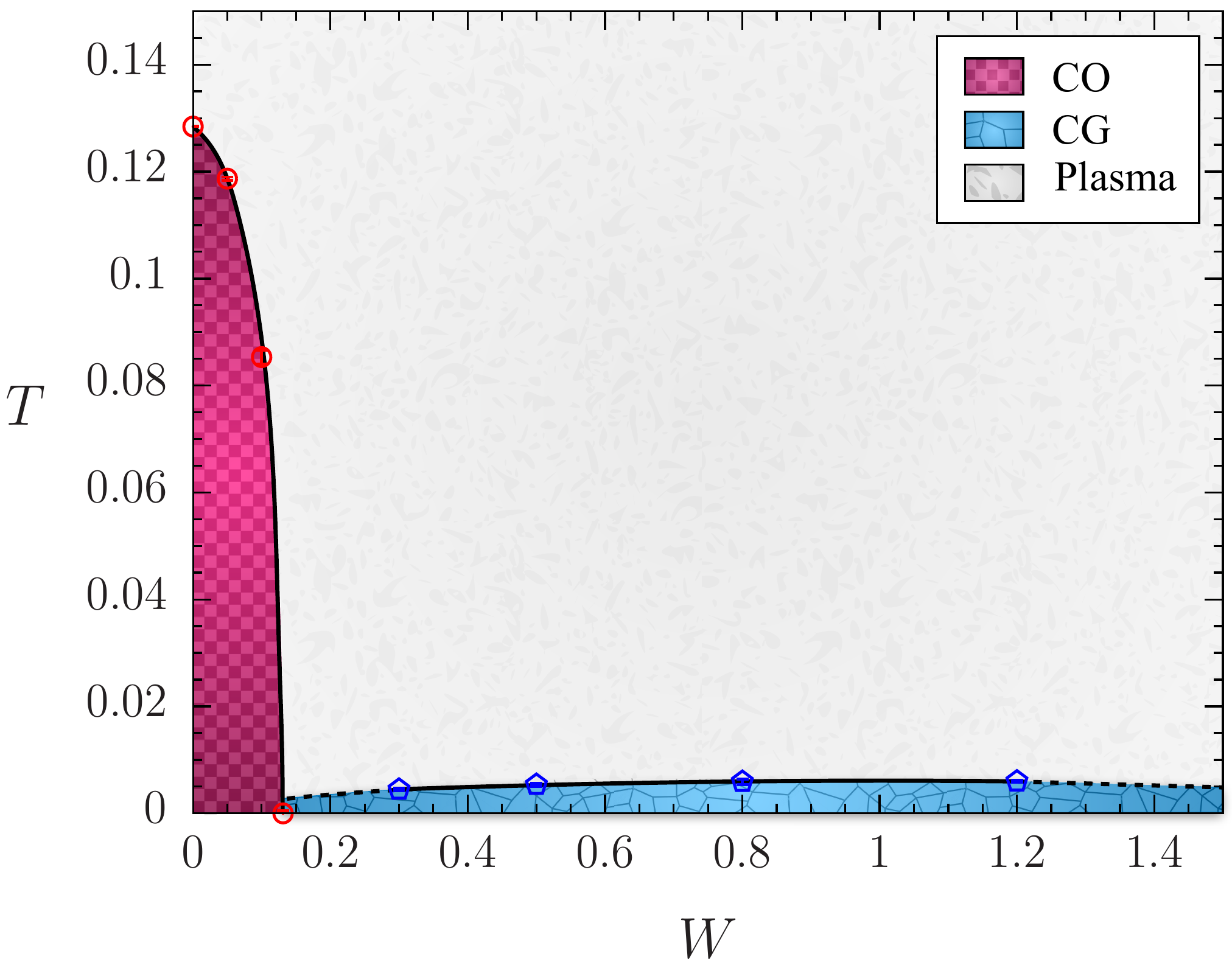}
\caption{
Phase diagram of the three-dimensional Coulomb glass model. There is a
charge order (CO) phase for $W \lesssim 0.131$ where electrons and holes
form a checkerboard-like crystal. For $W \gtrsim 0.131$ the system
undergoes a glassy transition into the Coulomb glass (CG) phase, albeit
at considerably lower temperatures than in the CO phase.  The dashed
lines indicate extrapolations where numerical simulations are not
available.}
\label{fig1:PhaseDiagram}
\end{figure}

The existence of disorder in strongly interacting electron
systems—which can be realized by introducing random impurities within
the material, e.g., a strongly doped semiconductor—plays a significant
role in understanding transport phenomena in imperfect materials and bad metals,
as well as in condensed matter in general.  When the density of
impurities is sufficiently large, electrons become localized via the
Anderson localization mechanism \cite{anderson:58} and the long-range
Coulomb interactions are no longer screened. This, in turn, leads to
the depletion of the single-particle density of states (DOS) near the
Fermi level, as first proposed by Pollak \cite{pollak:70} and Srinivasan
\cite{srinivasan:71}, thus forming a pseudogap.  Later, Efros and
Shklovskii \cite{efros:75} (ES) solidified this observation by
describing the mechanisms involved in the formation of this pseudogap.
The ES theory explains how the hopping (DC conductivity) within a
disordered insulating material is modified in the presence of a
pseudogap, also referred to as the ``Coulomb gap.'' Numerous analytic
studies have predicted,
\cite{gruenewald:82,pollak:92,vojta:93,pastor:99,pastor:02,dobrosavljevic:03,mueller:04,pankov:05,
lebanon:05,amir:08},
as well as experimental studies observed
\cite{monroe:87,benchorin:93,massey:95,ovadyahu:97,martinez-arizala:98,vaknin:00,
bogdanovich:02,vaknin:02,orlyanchik:04,romero:05,jaroszynski:06,grenet:07,ovadyahu:07,raicevic:08,raicevic:11},
the emergence of glassy properties in such disordered insulators,
leading to the so-called ``Coulomb glass'' (CG) phase.  Experimentally,
to date, none of the aforementioned studies have observed a true
thermodynamic transition into a glass phase but rather have found
evidence of nonequilibrium glassy dynamics, i.e., dynamic phenomena that
are suggestive of a glass phase, such as slow relaxation, aging, memory
effects, and alterations in the noise characteristics.  Theoretically,
more recent seminal mean-field studies by Pankov and Dobrosavljevi{\'c}
\cite{pankov:05}, as well as M{\"u}ller and Pankov \cite{mueller:07}
have shown that there exists a marginally stable glass phase within the
CG model whose transition temperature $T_{\rm c}$ decreases as $T_{\rm c}\sim
W^{-1/2}$ for large enough disorder strength $W$, and is closely related
to the formation of the Coulomb gap. Whether the results of the
mean-field approach can be readily generalized to lower space dimensions
is still uncertain. However, as we show in this work, the mean-field
results of Ref.~\cite{pankov:05} quantitatively agree with our numerical
simulations in the charge-ordered regime (see
Fig.~\ref{fig1:PhaseDiagram}) with similar values for the critical
disorder $W_{\rm c}$ where the charge-ordered phase is suppressed.
The critical temperatures $T_{\rm c}$ for the glassy phase , on the other hand,
are substantially smaller than in the mean-field predictions.  This, in turn, suggests that the
mean-field approach of Ref.~\cite{pankov:05} includes the fluctuations of
the uniform charge order collective modes, but not of the glassy
collective modes.

There have been multiple numerical studies that attempt to both
understand the DOS, as well as the nature of the transitions of the CG
model. In fact, there has even been some slight disagreement as to what
 the theoretical model to simulate should be with some arguing for
lattice disorder to introduce randomness into the model
\cite{grannan:93, vojta:94} and others suggesting that the disorder should be
introduced via random biases. Numerically, a Coulomb gap in agreement
with the ES theory has been observed in multiple studies. However, there
is no consensus in the vast numerical work
\cite{baranovskii:79,davies:82,davies:84,lee:88,moebius:92,grannan:93,li:94,
sarvestani:95,wappler:97,diaz-sanches:00,sandow:01,grempel:04,overlin:04,kolton:05,
glatz:07,goethe:09,surer:09,moebius:09b,palassini:12,rehn:16,goethe:18x}
on the existence of a thermodynamic transition into a glassy phase.
Nonequilibrium approaches suggest the existence of glassy behavior; 
however, thermodynamic simulations have failed to detect a clear
transition.

In this paper we investigate the phase diagram of the CG model using
Monte Carlo simulations in three spatial dimensions. For the
finite-temperature simulations we make use of the population annealing
Monte Carlo (PAMC) algorithm \cite{hukushima:03,machta:10,wang:15e,amey:18,barzegar:18}
which enables us to thermalize for a broad range of disorder values down
to unprecedented low temperatures previously inaccessible.  In addition,
we argue that the detection of a glass phase requires a four-replica
correlation length, as commonly used in spin-glass simulations in a
field \cite{young:04,katzgraber:05c}.  Our main result is shown in
Fig.~\ref{fig1:PhaseDiagram}. Consistently with previous numerical and
analytical studies \cite{malik:07,pankov:05,goethe:09} we find a charge
ordered (CO) phase for disorders lower than $W_{\rm c}=0.131(2)$ where
electrons and holes form a checkerboard-like crystal. This is in close
analogy with the classical Wigner crystal \cite{wigner:34} which happens
at low electron densities where the potential energy dominates the
kinetic energy resulting in an ordered arrangement of the charges. 
It should however be noted that at $W\!=\!0$ the lattice model, unlike in the
continuum case, is not a standard Wigner crystal \cite{pramudya:11}
because the system exhibits a pseudogap in the excitation spectrum
(unrelated to the Coulomb gap) prior to entering the charge-ordered phase.
For disorders larger than $W_{\rm c}$ we find strong evidence of a
thermodynamic glassy phase restricted to temperatures which are
approximately one order of magnitude smaller compared to the CO
temperature scales. This, in turn, suggests that, indeed, a
thermodynamic glassy phase can exist in experimental systems where
typically off-equilibrium measurements are performed. It also resolves
the long-standing controversy where numerical simulations were unable to
conclusively detect a thermodynamic glassy phase while mean-field theory
predicted such a phase.  We note that for the disorder strength values
studied, we are unable to discern a monotonic decrease in the critical
temperature, as suggested by mean-field theory.

The paper is structured as follows. In Sec.~\ref{section:2} we
introduce the CG model, followed by the details of the simulation in
Sec.~\ref{section:3}.  Section \ref{section:4} is dedicated to the
results of the study. Concluding remarks are presented in
Sec.~\ref{section:5}.

\section{Model}
\label{section:2}
The CG model in three spatial dimensions is described by the Hamiltonian
\begin{align}
\mathcal{H}=\frac{e^2}{2\kappa}\sum_{i\neq
  j}(n_i-\nu)\frac{1}{|\mathbf{r}_{ij}|}(n_j-\nu) +
  \sum_i{n_i\phi_i},
\label{charge-Hamiltonian}
\end{align}
where $\kappa=4\pi\epsilon_0$, $n_i\in \{0,1\}$, and $\nu$ is the
filling factor. The disorder $\phi_i$ is an on-site Gaussian random
potential, i.e., $\mathcal{P}(\phi_i)=\left(2\pi
W^2\right)^{-1/2}\exp\left(-{\phi_i^2}/{2W^2}\right)$. At half filling
($\nu=1/2$) the CG model can conveniently be mapped to a long-range spin
model via $s_i=(2n_i-1)$.  The Hamiltonian can be made dimensionless by
choosing the units such that $e^2/\kappa=1$ and $a=1$ in which $a$ is
the lattice spacing.  We thus simulate
\begin{align}
\mathcal{H}=\frac{1}{8}\sum_{i\neq j}\frac{s_is_j}{|\mathbf{r}_{ij}|} +
\frac{1}{2}\sum_i{s_i\phi_i},
\label{spin-Hamiltonian}
\end{align}
where $s_i \in \{\pm 1\}$ represent Ising spins.

\section{Simulation details}
\label{section:3}
In order to reduce the finite-size effects we use periodic boundary
conditions. Special care has to be taken to deal with the long-range
interactions. We make infinitely many periodic copies of each spin in
all spatial directions, such that each spin interacts with all other
spins infinitely many times. We use the Ewald summation technique
\cite{ewald:21g,deleeuw:80}, such that the double summation in
Eq.~\eqref{spin-Hamiltonian} can be written in the following way
\begin{align}
\frac{1}{2}\sum_{i=1}^{N}\sum_{j=1}^{N}s_is_j\left[f^{(1)}_{ij}+f^{(2)}_{ij}+f^{(3)}_{ij}+f^{(4)}_{ij}\right],\label{Ewald-sum}
\end{align}
where the terms $f_{ij}$ are defined as
\begin{align}
&f^{(1)}_{ij}=
  \frac{1}{4} \sideset{}{'}\sum_{\mathbf{n}}
  \frac{\textrm{erfc}\left(\alpha |\mathbf{r}_{ij}+\mathbf{n}L|\right)}{|\mathbf{r}_{ij}+\mathbf{n}L|},
  \label{short-range}\\
&f^{(2)}_{ij}=
  \frac{\pi}{N}\sum_{\mathbf{k}\neq{0}}
  \frac{\mathrm{e}^{-\mathbf{k}^2/4\alpha^2}}{\mathbf{k}^2}\cos(\mathbf{k}\mathbf{r}_{ij}),
  \label{long-range}\\
&f^{(3)}_{ij}=
  \frac{\pi}{3N}\mathbf{r}_i.\mathbf{r}_j,
  \label{dipole}\\[0.7em]
&f_{ij}^{(4)}=
  -\frac{\alpha}{2\sqrt{\pi}} \delta_{ij}.
  \label{self-energy}
\end{align}
Here, $\textrm{erfc}$ is the complimentary error function
\cite{abramowitz:64}, $\alpha$ is a regularization parameter, and
$\mathbf{k}={2\pi\mathbf{n}}/{L}$ is the reciprocal lattice momentum.
The vector index $\mathbf{n}$ in Eq.~\eqref{short-range} runs over the
lattice copies in all spatial directions and the prime indicates that
$\mathbf{n}=0$ is not taken into account in the sum when $i=j$.  For
numerical purposes, the real and reciprocal space summations, i.e.,
Eqs.~\eqref{short-range} and \eqref{long-range}, respectively, are
bounded by $|\mathbf{r}_{ij}+\mathbf{n}L|<r_c$ and $k<{2\pi n_c}/{L}$.
The parameters $\alpha$, $r_c$, and $n_c$ are tuned to ensure a stable
convergence of the sum. We find that $2<\alpha < 4$, $n_c\gtrsim{4L}$,
and $r_c=L/2$ are sufficient for the above purpose.

We use population annealing Monte Carlo (PAMC)
\cite{hukushima:03,machta:10,wang:15e,amey:18,barzegar:18} to thermalize the system down to
extremely low temperatures. In PAMC, similarly to simulated annealing (SA)
\cite{kirkpatrick:83}, the system is equilibrated towards a target
temperature starting from a high temperature following an annealing
schedule.  PAMC, however, outperforms SA by introducing many replicas of
the same system and thermalizing them in parallel. Each replica is
subjected to a series of Monte Carlo moves and the entire pool of
replicas is resampled according to an appropriate Boltzmann weight. This
ensures that the system is equilibrated according to the Gibbs
distribution at each temperature. For the simulations we use
particle-conserving dynamics to ensure that the lattice half filling is kept
constant, together with a hybrid temperature schedule linear in $\beta$
and linear in $T$ \cite{barzegar:18}.  We use the family entropy of
population annealing \cite{wang:15e} as an equilibration criterion. Hard
samples are re-simulated with a larger population size and number of
sweeps until the equilibration criterion is met. Note that we have
independently examined the accuracy of the results, as well as the
quality of thermalization for system sizes up to $L=8$ using parallel
tempering Monte Carlo \cite{hukushima:96}.  Both data from PAMC and
parallel tempering Monte Carlo agree within error bars.  We investigate
the phase diagram of the CG model using fixed values of the disorder
width, i.e., vertical cuts on the $W$–$T$ plane. Further details of the
simulation parameters can be found in Tables~\ref{simpar_co} and
\ref{simpar_cg} for the CO and CG phases, respectively.

\begin{table}[t!]
\caption{
PAMC simulation parameters used for the finite-temperature simulations
in the CO phase ($W \le 0.131$).  $L$ is the linear system size, $R_0$
is the initial population size, $M$ is the number of Metropolis sweeps, $T_0$
is the lowest temperature simulated, $N_T$ is the number of temperatures
and $N_{\rm sa}$ is the number of disorder realizations. Note that the
the values in the table vary slightly for different values of the
disorder $W$.}
\begin{tabular*}{\columnwidth}{@{\extracolsep{\fill}} l l l l l r}
\hline
\hline
$L$ & $R_0$ & $M$ & $T_0$ & $N_T$ & $N_{\rm sa}$\\
\hline
$4$ & $2\times{10^4}$ & $10$ & $0.05$ & $401$  & $5000$\\
$6$ & $5\times10^4$  & $10$ & $0.05$ & $601$ & $5000$\\
$8$ & $1\times{10^5}$  & $20$ & $0.05$ & $801$  & $2000$ \\
$10$ & $2\times{10^5}$  & $20$ & $0.05$ & $1001$  & $1000$\\
$12$ & $5\times10^5$  & $30$ & $0.05$ & $1201$ & $500$\\
\hline
\hline
\end{tabular*}
\label{simpar_co}
\end{table}

\begin{table}[b!]
\caption{
PAMC simulation parameters used for the finite-temperature simulations
in the CG phase ($W > 0.131$). For details see the caption of
Table~\ref{simpar_co}. Note that the the values in the table vary
slightly for different values of the disorder $W$.}
\begin{tabular*}{\columnwidth}{@{\extracolsep{\fill}} l l l l l r}
\hline
\hline
$L$ & $R_0$ & $M$ & $T_0$ & $N_T$ & $N_{\rm sa}$\\
\hline
$4$ & $2\times{10^4}$ & $20$ & $0.004$ & $401$  & $100000$\\
$6$ & $5\times10^4$  & $30$ & $0.004$  & $601$ & $50000$\\
$8$ & $1\times{10^5}$ & $40$ & $0.004$ & $801$  & $30000$\\
$10$ & $2\times{10^5}$ & $60$ & $0.004$ & $1001$  & $20000$\\
\hline
\hline
\end{tabular*}
\label{simpar_cg}
\end{table}

\section{Results}
\label{section:4}
\subsection{Charge-ordered phase}
\label{CO-phase}

\begin{figure}[th!]
\includegraphics[width=\columnwidth]{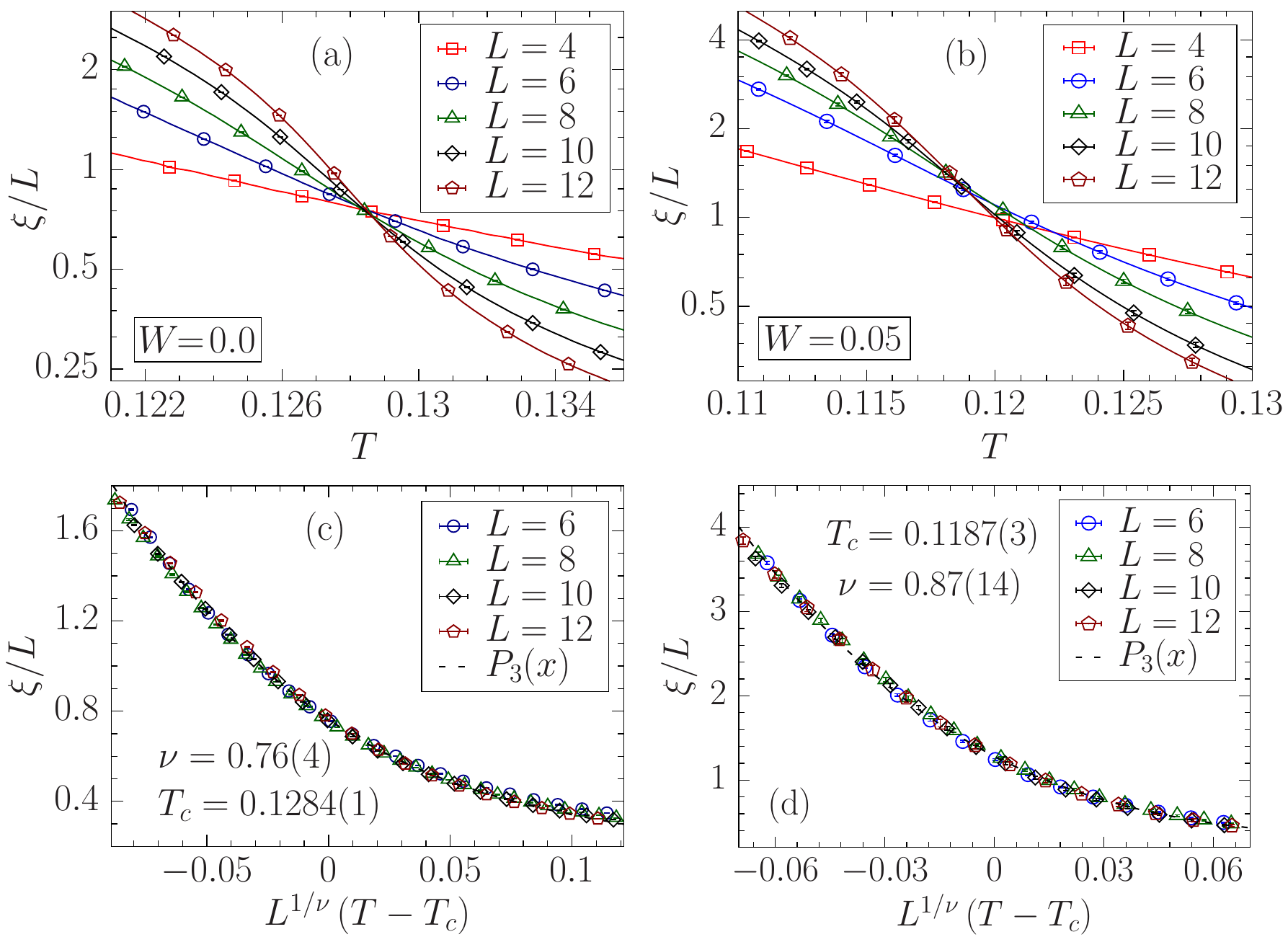}
\caption{Finite-size correlation length per system size $\xi/L$ versus
temperature $T$ for various disorder strengths. (a) No disorder, (b)
small disorder ($W=0.05$). In both cases we observe a crossing of the
data for different system sizes, suggesting a phase transition between a
disordered electron plasma and a CO phase.  (c), (d) Finite-size scaling
analysis used to determine the best estimates for the critical
temperature $T_{\rm c}$, as well as the critical exponent $\nu$ at the
aforementioned disorder values. Note that the smallest system size is
left out of the analysis for better accuracy. The transition temperature
$T_{\rm c}$ of the CO phase decreases as the disorder grows. }
\label{fig2:xi-CO}
\end{figure}

To characterize the CO phase, we measure the specific heat capacity
$c_{\rm v}=C_{\rm v}/N$ (only used to extract critical exponents, see
Appendix~\ref{App.B:FSS_results} for details), staggered magnetization
\begin{equation}
m_{\rm s}=\frac{1}{N}\sum_{i=1}^N\sigma_i ,
\end{equation}
where $\sigma_i=(-1)^{x_i+y_i+z_i}s_i$ and $N = L^3$ the number of spins,
as well as the disconnected and connected susceptibility
\begin{align}
\bar{\chi} &= N[\langle m_{\rm s}^2\rangle], \label{disconnected-susceptibility}\\
\chi &= N[\langle m_{\rm s}^2\rangle-\langle\left|m_{\rm s}\right|\rangle^2].\label{connected-susceptibility}
\end{align}
In addition, we measure the Binder ratio $g$ \cite{binder:81},
\begin{align}
&g=\frac{1}{2}\left(3-\frac{\left[\langle m_{\rm s}^4\rangle\right]}
  {\left[\langle m_{\rm s}^2\rangle\right]^2}\right),\label{Binder-ratio}
\end{align}
and the finite-size correlation length $\xi/L$
\cite{cooper:82,palassini:99b,ballesteros:00}, defined via
\begin{align}
&\xi=\frac{1}{2\sin\left(|\mathbf{k}_{\mathrm{min}}|/2\right)}
  \left(\frac{\chi(0)}{\chi(\mathbf{k}_{\mathrm{min}})}-1\right)^{1/2}
  ,\label{correlation-length}
\end{align}
where $\mathbf{k}_{\mathrm{min}}=(2\pi/L,0,0)$ is the smallest nonzero wave
vector and
\begin{equation}
\chi(\mathbf{k}) = \frac{1}{N}\sum_{ij} [\langle \sigma_i
\sigma_j \rangle] \exp(i \mathbf{k}.\mathbf{r}_{ij})
\end{equation}
is the Fourier transform of the susceptibility. Furthermore, $\langle
\cdots \rangle$ represents a thermal average and $[\cdots]$ is an average
over disorder. According to the scaling ansatz, in the
vicinity of a second-order phase transition temperature $T_{\rm c}$, any
dimensionless thermodynamic quantity such as the Binder ratio and the
finite-size correlation length divided by linear system size will be a
universal function of $x=L^{1/\nu}(T-T_{\rm c})$, i.e.,
$g=\tilde{F}_g(x)$ and $\xi/L=\tilde{F}_\xi(x)$, where $\nu$
is a critical exponent. Therefore, an effective way of probing a phase
transition is to search for a point where $g$ or $\xi/L$ data
intersect. Given the universality of the scaling functions
$\tilde{F}_g$ and $\tilde{F}_\xi$, if one plots $g$ or $\xi/L$
versus $x=L^{1/\nu}(T-T_{\rm c})$, the data for all system sizes must collapse
onto a common curve. Because we are dealing with temperatures close to
$T_{\rm c}$, we may approximate this universal curve by an appropriate mathematical
function such as a third-order polynomial $f(x) = P_3(x)$ in the case of $\xi/L$
or a complimentary error function $f(x) = \frac{1}{2}{\rm erfc}(x)$ when studying the Binder cumulant.
Hence, by fitting $f(x)$ to the data with $T_{\rm c}$ and $\nu$ as part of the fit parameters,
we are able to determine their best estimates.
\newpage
\onecolumngrid
\begin{table*}[t!]
\caption{Critical parameters of the plasma–CO phase transition
at different disorder values. The exponents, except for $\nu$, change with disorder.
Note that at $T=0$, the exponents $\alpha$ and $\gamma$ have been calculated in
a different way (see text in Appendix~\ref{App.B:FSS_results}).}
\label{table2:exponents-CO}
\begin{tabular*}{\textwidth}{@{\extracolsep{\fill}} l l l l l l l r}
\hline
\hline
Model & $W$ & $T_{\rm c}$ & $\nu$ & $\alpha/\nu$ & $\beta/\nu$ & $\bar\gamma/\nu$  & $\gamma/\nu$\\
\hline
CG & $0.000$ & $0.1284(1)$ & $0.76(4)$ & $0.550(2)$ & $0.42(1)$ & $2.41(1)$ & $2.05(2)$ \\
CG & $0.050$ & $ 0.1187(3)$ & $0.87(14)$ & $0.418(25)$ & $ 0.305(19)$ & $2.67(2)$ & $ 1.79(3)$ \\
CG & $0.131(2)$ & $ 0.000$ & $0.71(5)$ & $0.006(31)$ & $0.154(5)$ & $2.88(1)$ & $1.55(4)$\\
%RFIM & $2.270(4)$ & $0.000$ & $1.37(9)$ &  $-0.06(4)$ & $0.013(1)$ & $2.93(11)$ & $1.50(3)$ \\
\hline
\hline
\end{tabular*}
\vspace{-0.4cm}
\end{table*}
\twocolumngrid

The statistical error bars of the fit parameters are calculated by bootstrapping
over the disorder realizations. In Fig.~\ref{fig2:xi-CO} we show the simulation
data as well as the finite-size scaling (FSS) plots for $\xi/L$ at two
different disorder values. Crossings can clearly be observed which
signals a phase transition into the CO phase.
Simulating multiple values of $W$, we observe a phase transition between
a disordered electron plasma and a CO phase for $W<0.131(2)$, consistent
with previous studies \cite{malik:07,pankov:05,goethe:09}. The CO phase
is a checkerboard-like crystal \cite{wigner:34}, where electrons and
holes form a regular lattice as the potential energy dominates the
kinetic energy at low temperatures.

We have also conducted zero-temperature simulations using simulated
annealing to determine the zero-temperature critical disorder $W_{\rm c}$ that
separates the CO from the CG phase.  We average over $N_\text{sa}=2048$
different disorder realizations for disorders $W>0.10$ and
$N_\text{sa}=512$ for $W\leq 0.10$. Each disorder realization is
restarted at least at $20$ different initial random spin configurations
and at each temperature step equilibrated $N_ \text{eq}$ Monte Carlo
steps. If at least $15\%$ of the runs reach the same minimal energy
configuration, we assume that the chosen $N_\text{eq}$ was large enough
and that the reached configuration is likely the ground state. If less
than $15\%$ of the configurations reach the minimal state, we increase
$N_\text{eq}$ and re-run the simulation until the $15\%$ threshold is
achieved. For the largest simulated system size ($L=8$) and large
disorders, typical equilibration times are $N_\text{eq}=2^{27}$ Monte
Carlo sweeps.

\begin{figure}[b!]
\includegraphics[width=\columnwidth]{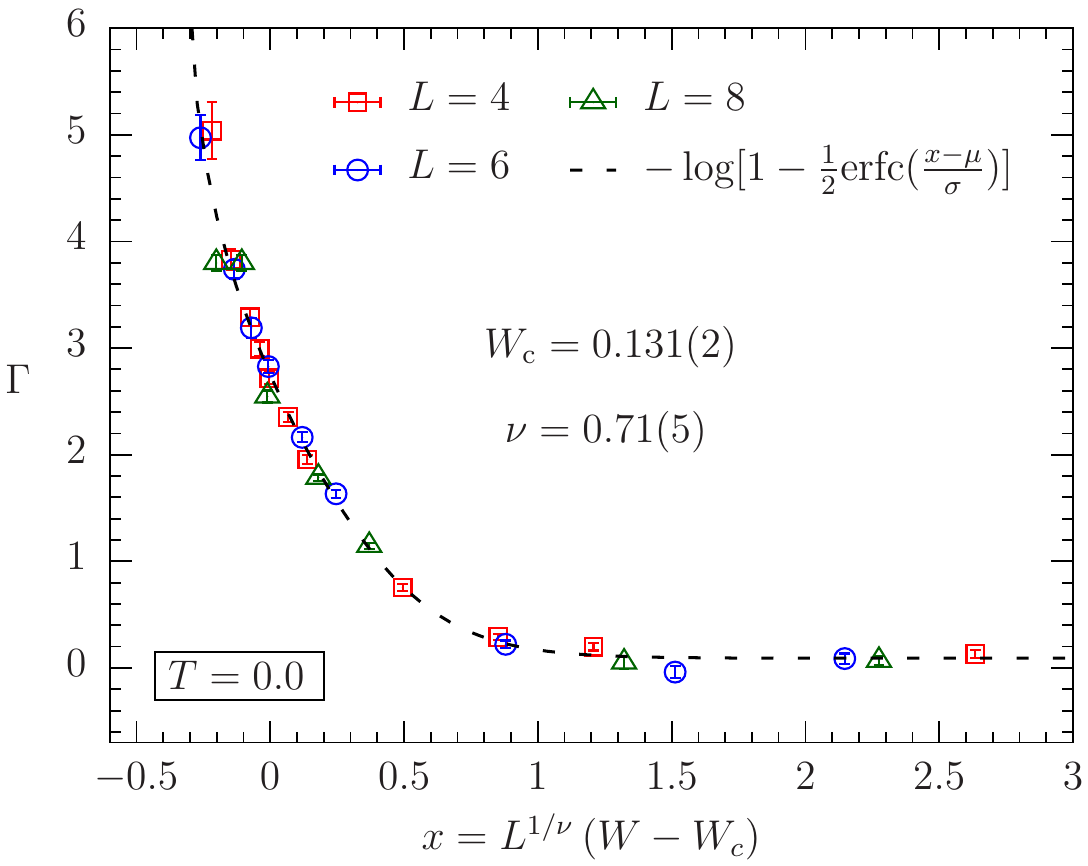}
\caption{Zero-temperature simulation results for the plasma–CO phase transition.
The quantity $\Gamma$ defined in Eq.~\eqref{Gamma} is used to perform a
finite-size scaling analysis. We conclude that the CO phase terminates
at $W_c=0.131(2)$. The statistical error bars are estimates by bootstrapping
over disorder instances. ${\rm erfc}(x)$ is the complimentary error function which is used to fit
the Binder ratio data (see text).}
\label{fig3:T=0}
\end{figure}

To estimate $W_{\rm c}$, we use the Binder ratio defined in Eq.~\eqref{Binder-ratio} which
by definition quickly approaches $1$ when $T\rightarrow 0$ within the CO phase.
Therefore, in order to retain a good resolution of a putative
transition, we use an alternative quantity $\Gamma$  which is
defined in the following way \cite{moebius:09b}:
\begin{align}
\Gamma=-\ln(1-g)\label{Gamma}.
\end{align}
Close to $W_{\rm c}$, we may assume the following finite-size
scaling behavior for $\Gamma$:
\begin{equation}
\Gamma=\tilde{F}_\Gamma\left[L^{1/\nu}(W-W_{\rm c})\right]
\end{equation}
As $g$ is restricted to $0\leq g \leq1$ with a step-function like shape, we may use
a complimentary error function $\frac{1}{2}{\rm erfc}(\frac{x-\mu}{\sigma})$
to represent the universal scaling function $\tilde{F}_\Gamma$ in which
$x=L^{1/\nu}(W-W_{\rm c})$ and $W_{\rm c}$, $\nu$, $\mu$,
$\sigma$ are the fit parameters. The fit is shown in Fig.~\ref{fig3:T=0}
where we obtain $W_{\rm c}=0.131\pm 0.002$ and $\nu = 0.71 \pm 0.05$.
In Table~\ref{table2:exponents-CO} (Appendix~\ref{App.B:FSS_results}) we list
the values of the critical exponents for the plasma–CO phase transition
for various disorder values $W$ after a comprehensive FSS analysis of
different observables. Note that we have used the methods developed in
Ref.~\cite{hartmann:01c} to compute the exponents $\alpha$ and $\gamma$
at $T\!=\!0$. An important observation one can promptly make is that the
exponents—except for $\nu$ which is universal—vary with disorder. This
can be attributed to the fact that the perturbations at large length
scales are contested between random field fluctuations which have static
nature and dynamic thermal fluctuations \cite{grinstein:76, bray:85,
fisher:86b}. At $W\!=0$, the perturbations are purely thermal, while at
$T\!=\!0$, the random field completely dominates.  At such large length
scales, the interactions within the charge ordered phase resemble the
random-field Ising model (RFIM)
\cite{middleton:02,fytas:13,fytas:13b,ahrens:13a} with short-range
bonds, namely, screening takes place.  This can be understood by
remembering that the dynamics of the system is constrained by charge
conservation.  In the spin language, excitations are no longer spin
flips but spin-pair flip-flops owing to the conservation of total
magnetization. For instance, one can create a local excitation while
preserving charge neutrality by moving a number of electrons out of a
subdomain in the CO phase. The excess energy of such a domain scales
like its surface, similarly to the short-range ferromagnetic Ising model.
It is worth mentioning that the Imry-Ma \cite{imry:75} picture gives a
lower critical dimension of $2$ for discrete spins with short-range
interactions.  Hence three-dimensional Ising spins, such as in the RFIM,
are stable to small random fields as we also find here.

Returning to the discussion of the critical exponents, we note that scaling relations such as
\begin{equation}
\gamma=\beta(\delta-1)=(2-\eta)\nu,
\end{equation}
as well as the modified  hyperscaling relation
\begin{equation}
(d- \theta)\nu=2-\alpha=2\beta+\gamma
\end{equation}
can be utilized to obtain estimates for the critical exponents $\eta$, $\theta$, and
$\delta$.  For instance, using the values in Table~\ref{table2:exponents-CO},
we see that $\eta(W = 0.0) = -0.05(2)$ and $\eta(W = 0.05) =
0.22(1)$. Near criticality, the correlation functions decay as a power
of distance, i.e.,  $G(\mathbf{x})\sim 1/|\mathbf{x}|^{d-2+\eta}$.  The
fact that the exponent $\eta$ is slightly negative for $W=0.0$ shows
that correlation between the spins remains in effect over a much longer
distance in the absence of disorder. Physically this is plausible, as
disorder tends to decorrelate the spins.

\subsection{Coulomb glass phase}
\label{CG-phase}
To examine the existence of a glassy phase in the CG model, we measure
the spin-glass correlation length defined in Eq.~\eqref{correlation-length},
however, for a spin-glass order parameter, namely
\begin{align}
&\xi_{\rm SG}=\frac{1} {2\sin\left(|\mathbf{k}_{\mathrm{min}}|/2\right)}
\left(\frac{\chi_{\rm SG}(0)}{\chi_{\rm SG}
(\mathbf{k}_{\mathrm{min}})}-1\right)^{1/2} .
\label{SG-correlation-Length}
\end{align}
Here, the spin-glass susceptibility $\chi_{\rm SG}$ has the following
definition \cite{ballesteros:00}:
\begin{align}
\hspace{-0.25cm}\chi_{\rm SG}(\mathbf{k}) = \frac{1}{N} \sum_{i=1}^N \sum_{j=1}^N\left[
  \left(\langle s_i s_j \rangle\!-\!\langle s_i \rangle \langle s_i \rangle
  \right)^2\right] \!\mathrm{e}^{i \mathbf{k}.( \mathbf{r}_i - \mathbf{r}_j)}. \label{SG-susceptibility}
\end{align}
It is important to note that $\langle s_i\rangle\neq 0$ because the
Hamiltonian [Eq.~\eqref{spin-Hamiltonian}] is not symmetric under global
spin flips. Therefore, at least four replicas are needed to compute the
connected correlation function in Eq.~\eqref{SG-susceptibility}.  We
start with the partition function of the system, using
Eq.~\eqref{spin-Hamiltonian}:
\begin{align}
Z=\sum_{\{s_i\}}\exp\left[-\beta\left(\frac{1}{8}\sum_{i\neq j}\frac{s_is_j}{|\mathbf{r}_{ij}|} +
\frac{1}{2}\sum_i{s_i\phi_i}\right)\right].
\label{partition-function}
\end{align}
We may now expresses any combination of the spin moments in terms of the
replicated spin variables $s_i^\alpha$ in the following way
\begin{align}
&\langle s_{1_1}\ldots s_{1_{k_1}}\rangle^{l_1}\dots\langle s_{m_1}\ldots s_{m_{k_m}}\rangle^{l_m}\nonumber\\
&=\,\frac{1}{Z^n}\sum_{\{s_i^\alpha\}}\mathrm{e}^{-\beta\!\!\sum\limits_{\alpha=1}^n\!\!\mathcal{H}[\{s_i^\alpha\}]}s_{1_1}^{1}
\ldots s_{1_{k_1}}^{1}\cdots \,s_{m_1}^{n}\ldots s_{m_{k_{m}}}^{n}\nonumber\\
&=\,\frac{1}{n!}\sum_{\alpha_1\ldots\alpha_n}^n\langle s_{1_1}^{\alpha_1}\ldots s_{1_{k_1}}^{\alpha_1}\cdots\,s_{m_1}^{\alpha_{n}}\ldots s_{m_{k_m}}^{\alpha_{n}}\rangle,\label{relicated-moments}
\end{align}
where $n=l_1+\cdots+l_m$ is the total number of replicas and replica
indices $\alpha_1$, $\ldots$ $\alpha_{n}$ are all distinct.  As a special
case, one can show
\begin{align}
\left(\langle s_i s_j \rangle - \langle s_i \rangle \langle s_j \rangle\right)^2
=&\,\frac{2}{4!}\sum_{\alpha,\beta}^4\langle s_i^\alpha s_j^\alpha s_i^\beta s_j^\beta\rangle\nonumber\\
&-\frac{2}{4!}\sum_{\alpha,\beta,\gamma}^4\langle s_i^\alpha s_j^\alpha s_i^\beta s_j^\gamma\rangle\nonumber\\
&+\frac{1}{4!}\sum_{\alpha,\beta,\gamma,\lambda}^4\langle s_i^\alpha s_i^\beta s_j^\gamma  s_j^\lambda\rangle.\label{replica-connected}
\end{align}

\begin{figure}[t!]
\includegraphics[width=0.95\columnwidth]{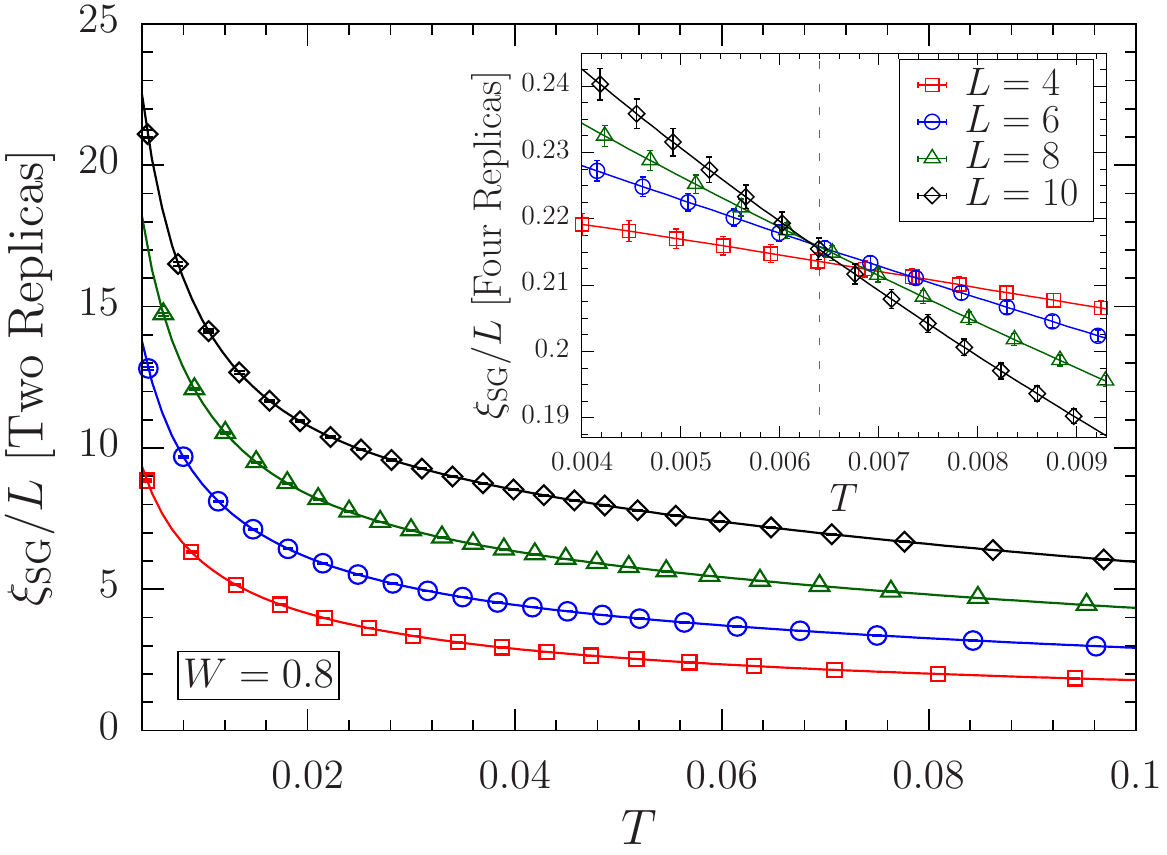}
\caption{
Spin-glass correlation length divided by system size $\xi_{\rm SG}/L$
calculated using two replicas at $W=0.8$ versus temperature $T$. No
crossing is observed down to very low temperatures. The inset shows the
same quantity using four replicas where a transition is clearly visible.
Here, data points for different system sizes cross approximately at the
temperature indicated by the dashed line. This suggests that in the
presence of external fields four-replica quantities need to be used to
characterize phase transitions in glassy systems.}
\label{2replica_vs_4replica}
\end{figure}

\begin{figure}[t!]
\includegraphics[width=\columnwidth]{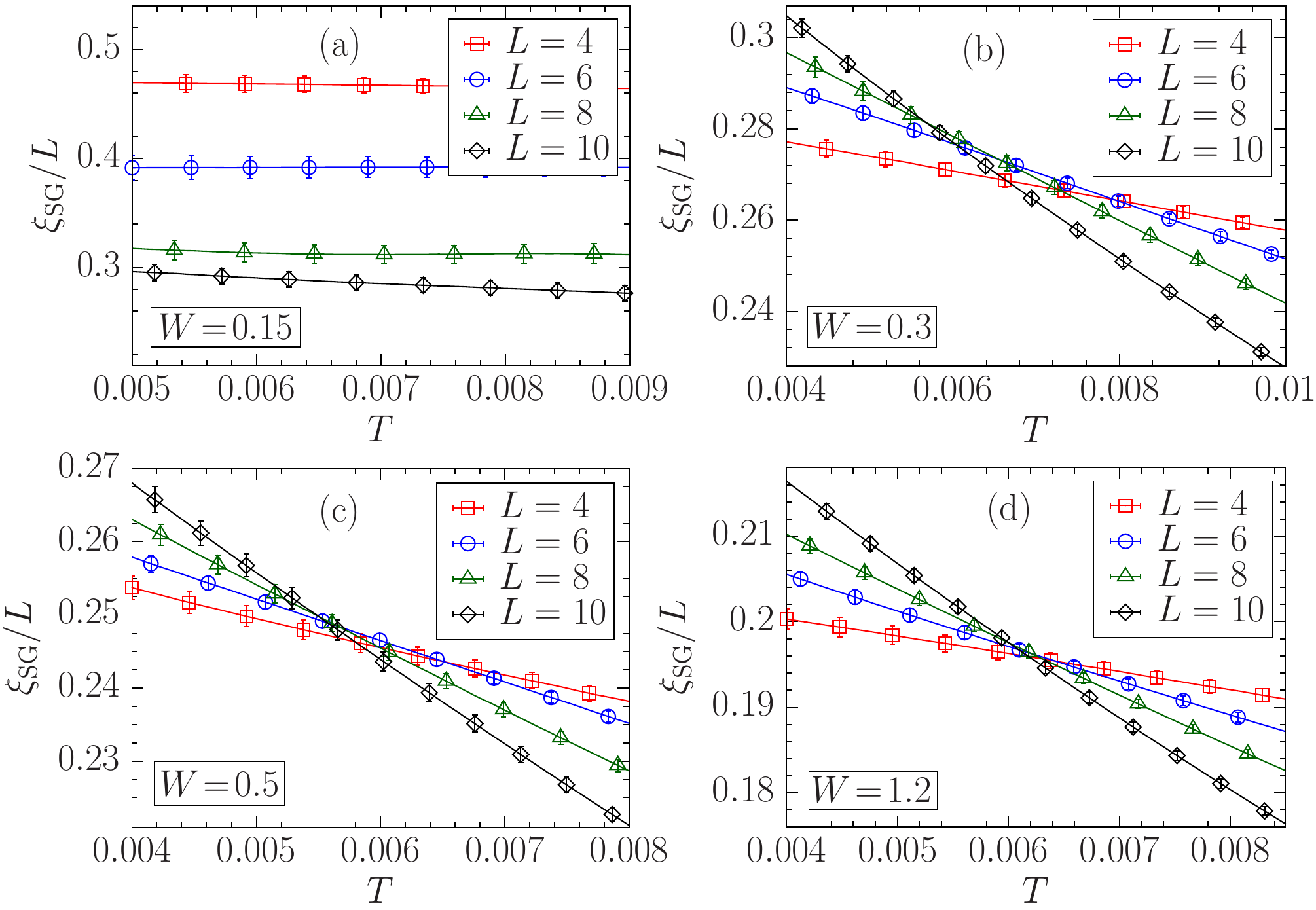}
\caption{
Spin-glass finite-size correlation length $\xi_{\rm SG}/L$ as a function
of temperature $T$ at various disorder strengths $W$.  (a) $W = 0.15$,
(b) $W = 0.30$, (c) $W = 0.50$, and (d) $W = 1.20$.  For  $W \gtrsim
0.15$ the data for different system sizes cross, indicating a plasma–CG
phase transition. Corrections to scaling must be considered to reliably
estimate the value of the critical temperature $T_{\rm c}$ (see text for details).}
\label{xi_SG_all}
\end{figure}

\noindent Using the above expression, the spin-glass susceptibility
[Eq.~\eqref{SG-susceptibility}] can be written in terms of the replica
overlaps as follows:
\begin{align}
\chi_{\rm SG}(\mathbf{k})=&\,\frac{N}{6}\sum_{\alpha<\beta}^4\left[
\langle q_{\alpha\beta}(\mathbf{k})q^*_{\alpha\beta}(\mathbf{k})\rangle\right]\nonumber\\
&-\frac{N}{6}\sum_{\alpha}^4\sum_{\beta<\gamma}^4\left[
\langle q_{\alpha\beta}(\mathbf{k})q^*_{\alpha\gamma}(\mathbf{k})\rangle\right]\nonumber\\
&+\frac{N}{3}\sum_{\alpha<\beta}^4\sum_{\gamma<\lambda}^4\left[
\langle q_{\alpha\beta}(\mathbf{k})q^*_{\gamma\lambda}(\mathbf{k})\rangle\right].\label{replica-susceptibility}
\end{align}
Once again, the indices $\alpha$, $\beta$, $\gamma$, and $\lambda$ must
be distinct.  Here, $q^*_{\alpha\beta}(\mathbf{k})$ represents the
complex conjugate of $q_{\alpha\beta}(\mathbf{k})$, and
$q_{\alpha\beta}(\mathbf{k})$ is the Fourier transformed spin overlap,
i.e.,
\begin{align}
q_{\alpha\beta}(\mathbf{k})=\frac{1}{N} \sum_{i=1}^N s_i^\alpha s_i^\beta\mathrm{e}^{i \mathbf{k}.\mathbf{r}_i }.\label{overlap}
\end{align}
To underline the significance of this matter, we have shown in
Fig.~\ref{2replica_vs_4replica} the spin-glass correlation length
calculated using two replicas, as has
been done in some previous numerical studies of the CG \cite{grannan:93, Bhandari:19}.
The inset shows the same quantity computed using four replicas.  While the
two-replica version of the finite-size correlation length shows no sign
of a CG transition, the four-replica expression captures the existence
of a phase transition into a glassy phase.

We have performed equilibrium simulations for $W\in\{0.15, 0.30, 0.50,
0.80,1.2\}$. In Fig.~\ref{xi_SG_all} we plot the four-replica spin-glass
correlation length as a function of temperature at selected disorder
values. Our results strongly suggest that there is a transition to a
glassy phase which persists for relatively large values of the disorder.
This is significant in the sense that it confirms the phase transition
via replica symmetry breaking as predicted by mean-field theory. The
nontriviality of our findings can be better understood if one juxtaposes
the CG case with that of finite-dimensional spin glasses lacking time
reversal symmetry due to an arbitrarily small external field where the
existence of de Almeida-Thouless \cite{almeida:78} transition, except
for a few rare cases \cite{banos:12,baity:14a}, has been ruled out by
numerous studies \cite{fisher:87,fisher:88,mattsson:95,jonsson:05,takayama:04,young:04}.
For the random-field Ising model the droplet picture of Fisher and Huse 
\cite{fisher:87,fisher:88} can be invoked to show the instability of the glass
phase to infinitesimal random fields. Yet, the CG model is different in two 
significant ways: typical compact domains are not charge neutral, and therefore
can not be flipped; and the long range of the interactions, while it does not affect 
the domain wall formation energy in the ordered phase, may be 
significant in the more complex domain formation of the glass phase.
It is worth emphasizing here that proper equilibration is key in observing a 
glassy phase in the CG simulations. For instance, in Fig.~\ref{poor-thermalization} of
Appendix~\ref{App.A:equilibration} we show an example of a simulation
where the crossing in the spin-glass correlation length is completely
masked due to insufficient thermalization.

Some corrections to scaling must be considered in the analysis in order
to estimate the position of the critical temperature and the values of
the critical exponents. In the vicinity of the critical temperature
$T_{\rm c}$ and to leading order in corrections to scaling, we may consider
the following FSS expressions for the spin-glass susceptibility
$\chi_{\rm SG}$ and the finite-size two-point correlation length divided
by the linear size of the system, $\xi_{\rm SG}/L$:
\begin{align}
&\chi_{\rm SG}\sim C_\chi L^{2-\eta}\left[1 + A_\chi L^{-\omega} + B_\chi L^{1/\nu}(T-T_{\rm c})\right],\label{chi_SG-scaling}\\
&\xi_{\rm SG}/L\sim C_\xi+A_\xi L^{-\omega} + B_\xi L^{1/\nu}(T-T_{\rm c}),
\label{xi_SG-scaling}
\end{align}
where $A_\chi$, $B_\chi$, $C_\chi$, $A_\xi$, $B_\xi$, and $C_\xi$  are constants. In order to find the critical temperature
$T_{\rm c}$ as well as the critical exponents $\nu$, $\eta$, $\omega$, we perform the following procedure.

\begin{enumerate}[label=(\roman*), wide, labelwidth=!, labelindent=0pt]
\item Estimation of  $T_{\rm c}$:
Given any pair of system sizes $(L_1,L_2)$ we have
\begin{align}
\label{mean L}
L_1=\bar{L}-\Delta L/2, \quad L_2=\bar{L}+\Delta L/2,
\end{align}
in which $\Delta L = L_2-L_1$ and $\bar{L} = (L_1+L_2)/2$.
Using Eq.~\eqref{xi_SG-scaling}, to the leading order in
$\Delta L/\bar{L}$  we find
\begin{align}
\frac{\xi_{\rm SG}(L_i,T)}{L_i} \sim\frac{ \xi_{\rm SG}(\bar{L},T)}{\bar{L}} -(-1)^i\frac{\Delta L}{2\bar{L}}\left[\omega A_\xi \bar{L}^{-\omega}\right.\nonumber\\
\left.- \frac{B_\xi}{\nu} \bar{L}^{1/\nu}(T-T_{\rm c})\right],\label{xi_SG_mean_L}
\end{align}
where the index $i$ can take values $i=1,2$.
One can now use Eq.~\eqref{xi_SG_mean_L} to determine the temperature $T^*(L_1,L_2)$ at which the
curves of $\xi_{\rm SG}/L$ cross; in other words, $\xi_{\rm SG}(L_1,T^*)/L_1=\xi_{\rm SG}(L_2,T^*)/L_2$ and
\begin{align}
T^*(L_1,L_2) \sim T_{\rm c} + \Theta_\xi\bar{L}^{-\omega-1/\nu}= T_{\rm c} + \Theta_\xi\bar{L}^{-\phi}.
\label{crossing-Temperature}
\end{align}
Here  $T_{\rm c}$ is the true critical temperature in the limit $L\rightarrow
\infty$ and $\Theta_\xi$ is a constant.
In Fig.~\ref{FSS_W=0.5}(a), we show the $T_{\rm c}$ estimate for the case $W=0.50$.
The best fit curve is obtained by minimizing the sum of the square of the residuals,
\begin{align}
\label{chisqr}
\chi^2 = \sum_{i=1}^N\left(T_i^*-T_{\rm c}-\Theta_\xi \bar{L}_i^{-\phi}\right)^2,
\end{align}
where $i$ runs over all pairs of linear system sizes. Now we vary $T_{\rm c}$, minimizing
$\chi^2$ along the way with respect to the remaining parameters. Since $\Theta_\xi$ appears
linearly in the model, it can be eliminated \cite{Lawton:71} to reduce the optimization
task to one free parameter, i.e., $\phi$:
\begin{align}
\label{reduction}
\left(\frac{\partial\chi^2}{\partial\Theta_\xi}\right)_{T_{\rm c}}\!\!\!= 0\,\,\Rightarrow \,\, \tilde{\Theta}_\xi(T_{\rm c}, \phi) = \frac{\sum\limits_{i=1}^N(T^*_i-T_{\rm c})\bar{L}_i^{-\phi}}{\sum\limits_{i=1}^N\bar{L}_i^{-2\phi}}.
\end{align}
Because there are five data points with three parameters in the original model,
we have two degrees of freedom. Therefore, the probability density function (PDF)
is proportional to $e^{-\chi^2/2}$.  

\newpage
\onecolumngrid
\begin{figure*}[th!]
\includegraphics[width=\textwidth]{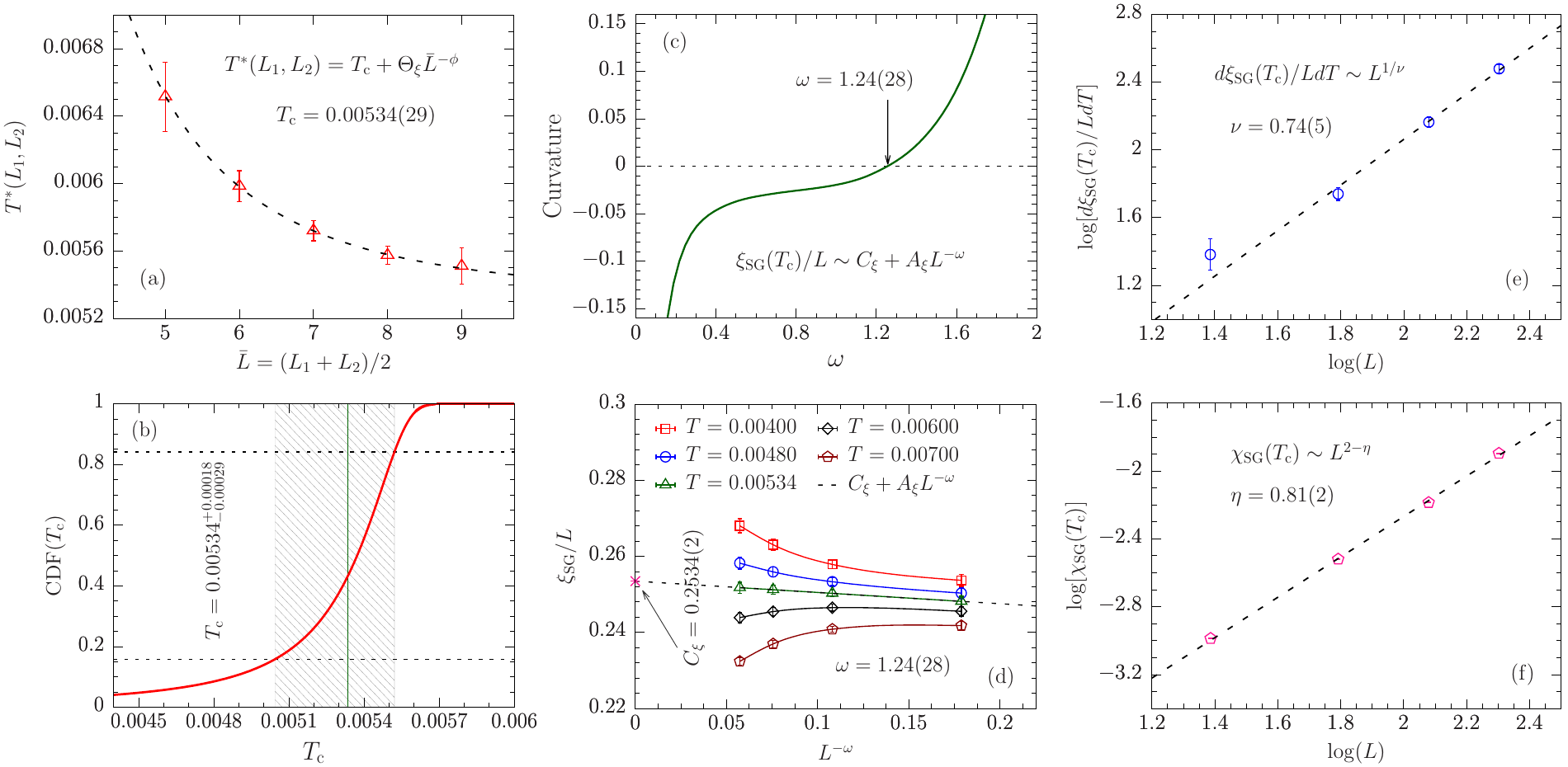}
\caption{
Process of estimating the critical exponents, as well as the critical
temperature $T_{\rm c}$ of the plasma–CG phase transition for $W=0.5$.  Other
values of $W$ are analyzed using the same procedure. (a) The
temperatures where  $\xi_{\rm SG}/L$ curves of different systems sizes
cross are used to determine the critical temperature $T_{\rm c}$. The crossing temperatures
decay toward the thermodynamic limit $T_{\rm c}$.
(b) The cumulative distribution function (CDF)
is constructed by minimizing  $\chi^2$  with respect to $\Theta_\xi$ and
$\phi$ while holding  $T_{\rm c}$ constant. The shaded region shows the $68\%$
confidence interval and the green vertical line indicates the best estimate of  $T_{\rm c}$.
(c) The value of $T_{\rm c}$ obtained in the
previous step is used to determine $\omega$.  At $T=T_{\rm c}$ and optimal $\omega$,
$\xi_{\rm SG}/L$ is linear as a functions of $L^{-\omega}$; i.e., it has zero curvature as demonstrated
in panel (d). (e) The critical exponent $\nu$ is estimated  using the derivative of
$\xi_{\rm SG}/L$  with respect to temperature which scales as $L^{1/\nu}$ when
evaluated at $T_{\rm c}$. Some deviations are evident
for the smallest system size. (f) The spin-glass susceptibility $\chi_{\rm SG}$ at $T=T_{\rm c}$
which scales as $L^{2-\eta}$ is used to determine the best estimate of the exponent $\eta$.}
\label{FSS_W=0.5}
\vspace{-0.5cm}
\end{figure*}
\twocolumngrid

To determine the confidence intervals,
we  calculate the cumulative distribution function (CDF) \cite{katzgraber:03f}:
\begin{align}
\label{CDF}
Q(T_{\rm c}) = \int^{T_{\rm c}}e^{-\frac{1}{2}\chi^2(T'_{\rm c})}dT'_{\rm c}.
\end{align}
As an example, in Fig.~\ref{FSS_W=0.5}(b) we have shown the $68\%$  confidence
 interval as well as the best estimate for the critical temperature.
\item Estimation of $\omega$:
From Eq.~\eqref{xi_SG-scaling} we observe that
\begin{align}
\label{xi_SG@Tc}
\xi_{\rm SG}(T_{\rm c})/L \sim C_\xi+A_\xi L^{-\omega}.
\end{align}
\begin{table}[b!]
\caption{Critical parameters of the plasma–CG phase transition for
various values of the disorder $W$. The exponent $\nu$ and $\omega$
are independent of $W$ within error bars highlighting their universality whereas the exponent
$\eta$ varies as  the disorder strength increases.}
\label{table3:exponents-CG}
\begin{tabular*}{\columnwidth}{@{\extracolsep{\fill}} l l l l r}
\hline
\hline
$W$ & $T_{\rm c}$ & $\nu$ & $\omega$ & $\eta$\\
\hline
$0.300$ & $0.00446(25)$ & $0.62(5)$ & $1.26(7)$ & $0.56(1)$ \\
$0.500$ & $0.00534(29)$ & $0.74(5)$ & $1.24(28)$ & $0.82(5)$\\
$0.800$ & $0.00590(56)$ & $0.64(2)$ & $1.28(20)$ & $ 0.97(5)$\\
$1.200$ & $0.00600(16)$ & $0.65(3)$ & $1.33(21)$ & $ 1.09(1)$\\
\hline
\hline
\end{tabular*}
\end{table}
Thus, using the best estimate of $T_{\rm c}$ from the previous step, we expect the
data points of $\xi_{\rm SG}(T_{\rm c})/L$ as a function of $L^{-\omega}$
to follow a straight line when $\omega$ is chosen correctly. We can therefore vary $\omega$ and measure the curvature
until it vanishes at the optimal value. We have demonstrated this in Figs.~\ref{FSS_W=0.5}(c)
and \ref{FSS_W=0.5}(d). Note that the error bar for $\omega$ is calculated using the bootstrap method.
\item Estimation of $\nu$ and $\eta$:
It is straightforward to show from Eqs.~\eqref{chi_SG-scaling} and \eqref{xi_SG-scaling} that to the
leading order in corrections,
\begin{align}
&\chi_{\rm SG}(T_{\rm c})= C_\chi L^{2-\eta}(1+A_\chi L^{-\omega}),\\
&\frac{d}{dT}(\xi_{\rm SG}/L)(T_{\rm c}) = B_\xi L^{1/\nu}(1+D_\xi L^{-\omega}),
\end{align}
in which the best estimates obtained for $T_{\rm c}$ and $\omega$
are used. We see that the above quantities simply scale as $\chi_{\rm SG}(T_{\rm c})\sim L^{2-\eta}$ and
$\frac{d}{dT}(\xi_{\rm SG}/L)(T_{\rm c}) \sim L^{1/\nu}$ for large enough $L$. 
Therefore, a linear fit in logarithmic scale will yield the exponents $\nu$ and $\omega$. 
This is shown in Figs.~\ref{FSS_W=0.5}(e) and \ref{FSS_W=0.5}(f), respectively.

The above procedure has been repeated for all
other values of the disorder $W$. The results are summarized
in Table~\ref{table3:exponents-CG} of Appendix~\ref{App.B:FSS_results}.
We observe that within the error bars, the critical exponents $\nu$
and $\omega$ are robust to disorder which underlines the universality of these exponents.
Nevertheless, larger system sizes—currently not accessible via simulation—would be needed to
conclusively determine the universality class of the model.  The fact that we observe stronger
corrections to scaling for smaller disorder shows that the energy landscape is
rougher due to competing interactions where finite-size effects are accentuated.
 For larger values of $W$, on the other hand, the system becomes easier to thermalize as the disorder
dominates the electrostatic interactions.
\end{enumerate}

\vspace{-0.3cm}

\section{Conclusion}
\label{section:5}
We have shown that, using the four-replica expressions for the
commonly-used observables, the CG model displays a transition into a
glassy phase  for the studied system sizes, provided that large enough
disorder and sufficiently low temperatures are used in the simulations
(see Fig.~\ref{fig1:PhaseDiagram} for the complete phase diagram of the
model). Previous numerical studies—including a work \cite{surer:09} by a
subset of us—have failed to observe the glassy phase.
In this study, we are able to present strong numerical
evidence for the validity of the mean-field results in three space
dimensions, which predicts transition to a glassy phase at large
disorder via replica symmetry breaking. Moreover, we corroborate
the results of previous studies for the low-disorder regime where a CO phase,
similar to the ferromagnetic phase in the RFIM, is observed. 
Interestingly, for large disorder values, the CG and
the RFIM are different, as the RFIM does not exhibit a transition into a
glassy phase (see, for example, Ref.~\cite{ahrens:13a} and references
therein).  A possible reason is the combination of the constrained
dynamics (magnetization-conserving dynamics) and the long-range Coulomb
interactions not present in the RFIM. These two factors can increase
frustration such that a glassy phase can emerge.  Our findings open the
possibility of describing electron glasses through an effective CG model
both theoretically and numerically.  Because most of the electron glass
experiments are performed in two-dimensional materials, it would be
desirable to investigate these results in two-dimensional models.  Our
preliminary results in two space dimensions show no sign of a glass
phase.

\vspace{-0.1cm}

\begin{acknowledgments}
The authors thank Vlad Dobrosavljevi{\'c}, Arnulf M{\"obius}, Wenlong
Wang, and A.~Peter Young for insights and useful discussions. We also
thank Darryl C.~Jacob for assistance with the simulations.  We thank
the National Science Foundation (Grant No.~DMR-1151387) for financial
support, Texas A\&M University for access to HPC resources (Ada and
Terra clusters), Ben Gurion University of the Negev for access to
their HPC resources, and Michael Lublinski for sharing with us his
CPU time. This work is supported in part by the Office of the Director
of National Intelligence (ODNI), Intelligence Advanced Research Projects
Activity (IARPA), via MIT Lincoln Laboratory Air Force Contract No.
FA8721-05-C-0002. The views and conclusions contained herein are those
of the authors and should not be interpreted as necessarily representing
the official policies or endorsements, either expressed or implied, of
ODNI, IARPA, or the U.S. Government. The U.S.  Government is authorized
to reproduce and distribute reprints for Governmental purpose
notwithstanding any copyright annotation thereon.

\end{acknowledgments}

\begin{appendix}

\section{Equilibration}
\label{App.A:equilibration}

In this appendix, we outline the steps taken to guarantee
thermalization. The data for this work are predominantly generated using
population annealing Monte Carlo (PAMC). In order to ensure that the
states sampled by a Monte Carlo simulation are in fact in thermodynamic
equilibrium, i.e., weighted according to the Boltzmann distribution, one
needs to strive against bias by controlling the systematic errors
intrinsic to the algorithm due to the finite population size.

\begin{figure}[t]
\includegraphics[width=\columnwidth]{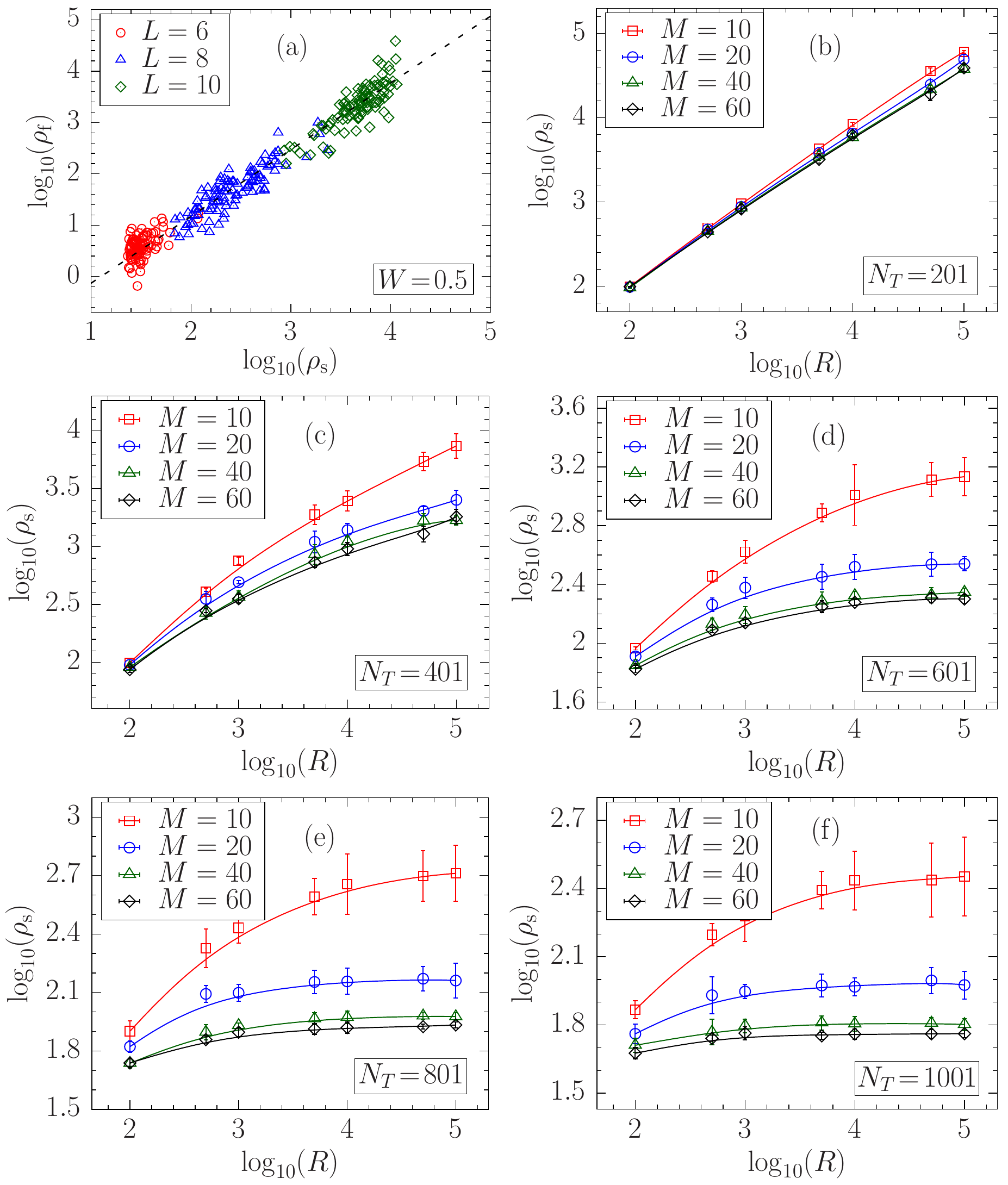}
\caption{
Equilibration of a PAMC simulation. (a) Equilibration
population size $\rho_{\rm f}$ versus entropic family size $\rho_{\rm s}$
for a CG simulations at $W=0.5$. $100$ instances for each system size
have been studied. Evidently, $\rho_{\rm s}$ is greatly correlated to
$\rho_{\rm f}$ which controls the systematic errors in thermodynamic
quantities.  Because $\rho_{\rm f}$ is computationally expensive to
measure, one may instead use $\rho_{\rm s}$ as the measure of
thermalization. (b)–(f) $\rho_{\rm s}$ versus the
population size $R$ for system size $L=8$ at various number of temperatures $N_T$ and
Metropolis sweeps $M$. When  $\rho_{\rm s}$ converges, the system is
guaranteed to be in thermal equilibrium. As seen from the plots,
convergence is achieved faster as the number of temperatures and sweeps
is increased. However, for extremely large values of $N_T$ and $M$,
marginal improvement in equilibration is gained at the cost of extended
run time of the simulation.
}
\label{optimization}
\end{figure}

Fortunately, PAMC offers a convenient way to study and tune the
systematic errors to a desired accuracy. It can be shown
\cite{wang:15e} that the systematic errors in a PAMC simulation are
directly proportional to the {\it equilibration population size}
$\rho_{\rm f}$ which has the following definition:
\begin{align}
\rho_{\rm f}=\lim_{R\rightarrow\infty}\!R\,{\rm var}(\beta F) .
\end{align}

\begin{figure}[t!]
\includegraphics[width=\columnwidth]{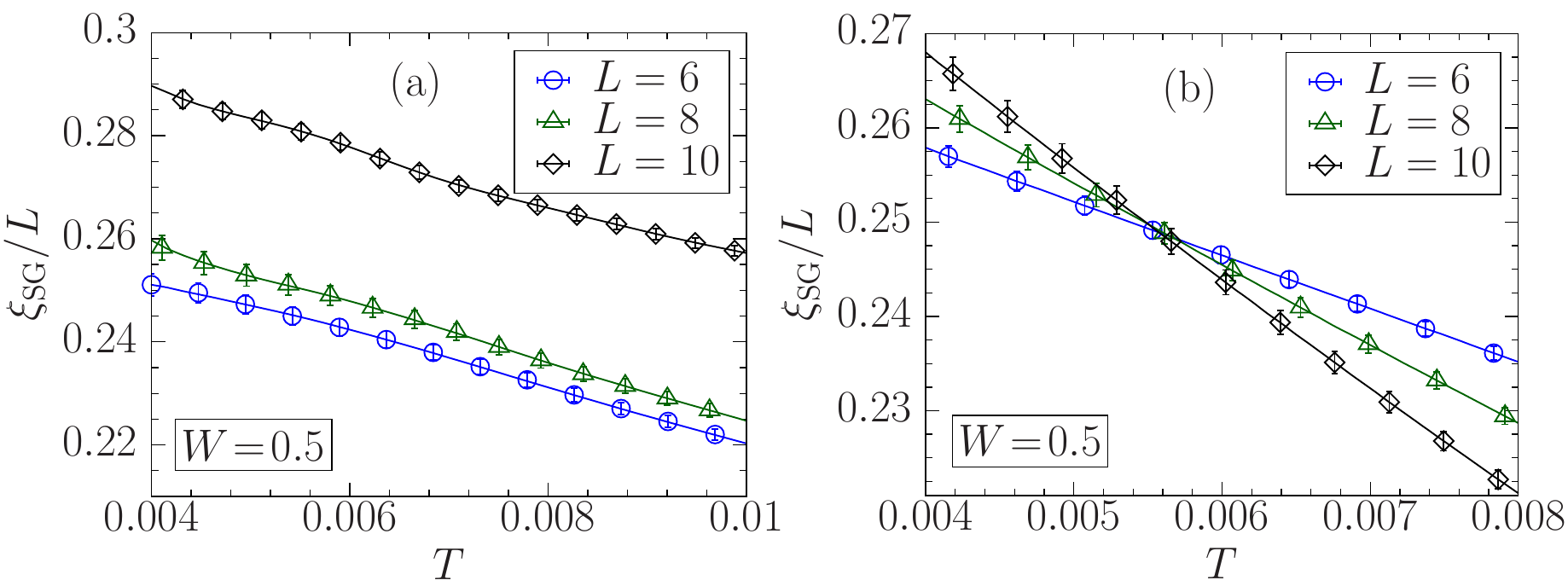}
\caption{
Importance of proper thermalization in observing a CG phase transition.
Panel (a) shows a simulation where some instances have not reached
thermal equilibrium whereas panel (b) illustrates the same simulation in
which all of the instances have been thoroughly thermalized.
}
\label{poor-thermalization}
\end{figure}
Here, $R$ is the population size and $F$ is the free energy.  $\rho_{\rm
f}$ is an extensive quantity defined at the thermodynamic limit although
in reality it converges at a large but finite $R$. Because $\rho_{\rm
f}$ is computationally expensive to measure as it requires multiple
independent runs, one may alternatively study the {\it entropic family
size} $\rho_{\rm s}$ defined as
\begin{align}
\rho_{\rm s}=\lim_{R\rightarrow\infty}\!R\,e^{-S_{\rm f}},
\end{align}
where $S_{\rm f}$ is the family entropy of PAMC. As shown in of
Fig.~\ref{optimization}(a), $\rho_{\rm s}$ is well correlated with
$\rho_{\rm f}$ which is why we can reliably use $\rho_{\rm s}$ as the
measure of equilibration.
$\rho_{\rm s}$ similarly to $\rho_{\rm f}$
converges at a finite $R$. The population size at which the convergence
is achieved is a function of the number of temperatures $N_T$ as well as
the number of Metropolis sweeps $M$. Optimization of PAMC is studied in
great detail in the context of spin glasses \cite{barzegar:18,amey:18}
much of which can be carried over to the CG simulations.
As an example we show in Figs.~\ref{optimization}(b)–\ref{optimization}(f) how we
choose the optimal values of the PAMC parameters. We observe that the
convergence of $\rho_{\rm s}$ is attained faster as the number of
temperatures and sweeps is increased. However, beyond a certain point,
any further increase solely prolongs the simulation time while
contributing negligibly to lowering the convergent value of  $\rho_{\rm
s}$. A good rule of thumb for checking thermalization,
as seen in Fig.~\ref{optimization}, is that $\rho_{\rm s}$ and as a result
$\rho_{\rm f}$ converges when $\rho_{\rm s}/R=\exp({-S_{\rm f}}) <  0.01$.
We ensure that the above criterion is met for every instance that we have studied.
This matter has been investigated thoroughly in Ref.~\cite{wang:15e}.
It is worth mentioning here that proper equilibration is crucial in
observing phase transitions, especially in subtle cases like the CG
model. We have illustrated this matter in
Fig.~\ref{poor-thermalization}. Figure~\ref{poor-thermalization}(a)
shows a simulation where the system has been poorly thermalized
in which $\rho_{\rm s}/R\sim0.1$ on average across the studied instances.
By contrast in Fig.~\ref{poor-thermalization}(b) the same simulation is done with
careful equilibration; that is to say, the criterion $\rho_{\rm s}/R <  0.01$
is strictly enforced for every instance. It is clear that the observation of a crossing is
contingent upon ensuring that every instance has reached thermal
equilibrium. This, in turn, could explain why simulations using parallel
tempering Monte Carlo, e.g., Ref.~\cite{goethe:09}, see no sign of a
transition.

\section{Finite-size scaling results}
\label{App.B:FSS_results}

\begin{figure}[t!]
\includegraphics[width=0.95\columnwidth]{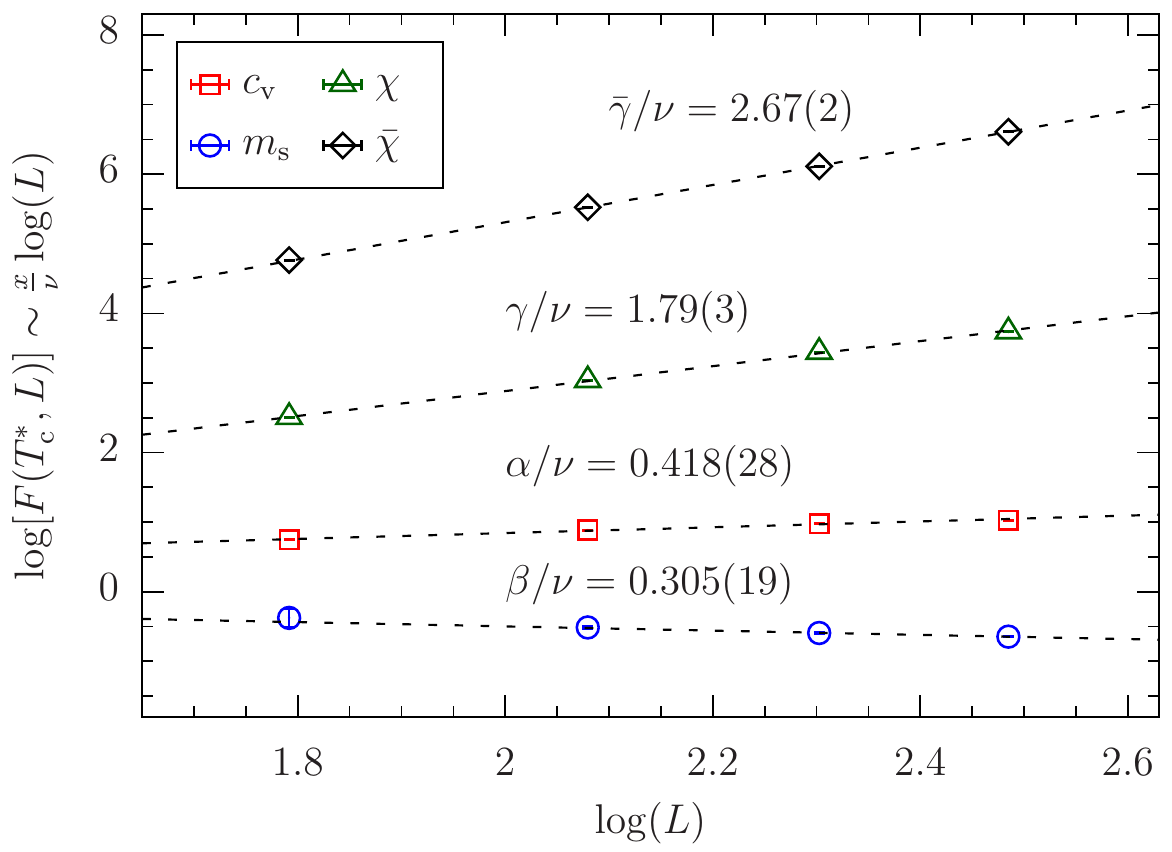}
\caption{
Finite-size scaling analysis for the plasma–CO phase transition at $W\!=\!0.05$.
The peak values of the specific heat capacity $c_{\rm v}$, connected and disconnected
susceptibilities $\chi$ and $\bar{\chi}$, as well as the inflection point value of the staggered magnetization
are used to estimate the critical exponents $\alpha$, $\beta$, $\gamma$, and $\bar{\gamma}$, respectively.
According to Eqs.~\eqref{CV-Ms-scaling} and \eqref{chi-chi_bar-scaling}, the above quantities scale as a power law in the
linear system size $L$ as clearly seen from the figure.
}
\label{peak_scaling}
\end{figure}

In this appendix we list the estimates for the critical parameters of
the plasma–CO, as well as the plasma–CG phase transitions. Because the
CO phase is essentially an antiferromagnetic phase in the spin language,
multiple critical exponents such as $\nu$, $\alpha$, $\beta$, and
$\gamma$ can be measured numerically. We have estimated these quantities
using FSS techniques, specifically by a FSS collapse of the data for
different system sizes onto a low-order polynomial, as explained in the
main text. To estimate the exponent $\nu$ we have used the finite-size
correlation length per linear system size $\xi/L$
[Eq.~\eqref{correlation-length}]. Because this is a dimensionless
quantity, in the vicinity of the critical point it scales as

\begin{align}
\xi/L&={F}_\xi\left[L^{1/\nu}(T-T_{\rm c})\right].
\label{Gamma-scaling}
\end{align}
Other critical exponents such as $\alpha$, $\gamma$, and $\beta$ can be
estimated by performing a FSS analysis using the peak values of the specific
heat $c_{\rm v}=C_{\rm v}/N$, connected susceptibility $\chi$, and
the disconnected susceptibility $\bar{\chi}$ as well as the inflection point value of the
staggered magnetization $m_{\rm s}$  which scale as following:
\begin{align}
&c_{\rm v}^{\rm max}\sim L^{\alpha/\nu}, \quad m_{\rm s}^{\rm inflect} \sim L^{-\beta/\nu}.\label{CV-Ms-scaling}\\
&\chi^{\rm max}\sim L^{\gamma/\nu}, \quad \bar{\chi}^{\rm max}\sim L^{\bar{\gamma}/\nu}.\label{chi-chi_bar-scaling}
\end{align}
As we can see in Fig.~\ref{peak_scaling} the above scaling behaviors are very well satisfied.
The best estimates of the critical parameters for various values of the disorder are listed
in Table~\ref{table2:exponents-CO}. Note that with the exception of the universal exponent $\nu$,
other critical exponents vary with disorder which can be due to the trade-off between
 large-scale thermal and random-field fluctuations.
Because at $T\!=\!0$  the system has settled in the ground state, one cannot use thermal sampling to measure
the variance of energy  and staggered magnetization which are proportional to
the heat capacity and susceptibility, respectively. Instead, we have used the techniques
developed by Hartmann and Young in Ref.~\cite{hartmann:01c}.

For the plasma–CG transition we have calculated the critical exponents
$\nu$ and $\eta$, as well as the correction to scaling exponent $\omega$,
using the procedure explained in Sec.~\ref{CG-phase}. Table~
\ref{table3:exponents-CG} lists the estimates of the critical
parameters.  Within the error bars, the exponents $\nu$ and $\omega$ are
independent of  disorder, whereas $\eta$  changes as the
disorder strength increases.

\end{appendix}

\bibliography{refs}

\end{document}